\documentclass[journal,final,twoside]{IEEEtran}
\usepackage{cite}

\ifCLASSINFOpdf
   \usepackage[pdftex]{graphicx}
   \graphicspath{{./figures/}{./pdf/}{./jpeg/}}
   \DeclareGraphicsExtensions{.pdf,.jpeg,.png}
\else
   \usepackage[dvips]{graphicx}
   \graphicspath{{./eps/}}
   \DeclareGraphicsExtensions{.eps}
\fi
\usepackage{algorithm}
\usepackage{algpseudocode}

\usepackage[cmex10]{amsmath}
\usepackage{amssymb}

\usepackage[caption=false,font=footnotesize]{subfig}
\usepackage{stfloats}

\usepackage{multirow}

\usepackage{footnote}

\usepackage{threeparttable}

\renewcommand{\vec}[1]{\mathbf{#1}}

\newcommand{\mtx}[1]{\mathbf{#1}}

\newcommand{\integerSequence}[1]{ \lbrack \lbrack {#1} \rbrack \rbrack }

\newcommand{\meanDistortion}[1]{\text{mean}^{\left({#1}\right)}_D}
\newcommand{\varDistortion}[1]{\text{var}^{\left({#1}\right)}_D}
\newcommand{\stdDistortion}[1]{\text{std}^{\left({#1}\right)}_D}

\begin{document}
%
\title{Benefiting from Duplicates of Compressed Data: \\ Shift-Based Holographic Compression of Images}

%
%
%

\author{Yehuda Dar and Alfred M. Bruckstein 
\\
\thanks{The authors are with the Department of Computer Science, Technion, Israel. E-mail addresses: \{ydar,~freddy\}@cs.technion.ac.il.}
}

%
%

\markboth{}%
{~}
%



\maketitle

\begin{abstract}
 
Storage systems often rely on multiple copies of the same compressed data, enabling recovery in case of binary data errors, of course, at the expense of a higher storage cost. In this paper we show that a wiser method of duplication entails great potential benefits for data types tolerating approximate representations, like images and videos.
We propose a method to produce a set of distinct compressed representations for a given signal, such that any subset of them allows reconstruction of the signal at a quality depending only on the number of compressed representations utilized. Essentially, we implement the holographic representation idea, where all the representations are equally important in refining the reconstruction. 
Here we propose to exploit the shift sensitivity of common compression processes and generate holographic representations via compression of various shifts of the signal. 
Two implementations for the idea, based on standard compression methods, are presented: the first is a simple, optimization-free design. The second approach originates in a challenging rate-distortion optimization, mitigated by the alternating direction method of multipliers (ADMM), leading to a process of repeatedly applying standard compression techniques. 
Evaluation of the approach, in conjunction with the JPEG2000 image compression standard, shows the effectiveness of the optimization in providing compressed holographic representations that, by means of an elementary reconstruction process, enable impressive gains of several dBs in PSNR over exact duplications.
 
\end{abstract}


~\\~

%
\IEEEpeerreviewmaketitle

\section{Introduction}
\label{sec:Introduction}

Any digital system involving storage or transmission of signals (e.g., images, videos and other multimedia data) fundamentally relies on lossy compression processes to meet storage-space or transmission bandwidth limitations, incurring acceptable reductions in the eventual recovered signal quality. 
Contemporary storage and content-distribution services implement processes where a binary compressed representation of a particular signal is exactly duplicated for the purpose of storage reliability, or for delivery to multiple users in a network. 
Clearly, subsequent access to several identical copies of the compressed signal cannot provide a reconstruction quality better than that achieved using a single copy. Hence, there is an inefficiency in the joint bit-cost of several copies versus the reconstruction quality they provide together. 
In this paper we address this type of inefficiency, as will be explained next.

Holographic representations \cite{bruckstein1998holographic,bruckstein2000holographic,bruckstein2018holographic} of a signal are a set of data packets designed so that its subsets enable signal approximation at a quality depending only on the number of packets utilized, and independent on the particular packets included in the subset. 
The holographic representations concept is closely related to the multiple description coding approach (see, e.g., \cite{goyal2001multiple,servetto2000multiple,jiang1998multiple}) as, indeed, both methods aim at reconstruction refinement when increasing the size of the subset of packets used for approximation. However, the two approaches differ in the following aspect: when using holographic representations, increasing the number of packets used for approximation leads to a quality gain (approximately) independent of the particular packets added at the expense of considerable higher bit-cost. In contrast, in multiple description coding, adding various packets may lead to considerably different quality gains due to serious concerns about keeping the bit-cost as low as possible \cite{goyal2001multiple,servetto2000multiple}. 
Inherently, the property of holographic representations implies that some amount of redundancy remains among the packets and, therefore, the packet bit-costs may be higher than in the multiple description coding approach. Nevertheless, the special properties of the holographic representations can significantly contribute to storage system designs.

In the context of storage systems, the holographic representations are intended for improving settings where several identical copies of compressed data are stored and their individual usefulness for recovery is more important than achieving the best possible reduction in their joint bit-cost. A prevalent case where single copy usefulness in reconstruction is crucial is in duplication-based reliable storage systems, where multiple identical versions of the data are stored for enabling recovery in case of errors in the binary form of the data. 
This approach is realized by the Redundant Array of Independent Disks (RAID) \cite{patterson1988case} data storage technology in mirroring-based settings.

In this paper, we focus on signals like audio, images and videos, commonly represented and processed in conjunction with lossy compression. Using the principles of holographic representations, we establish a methodology to store a signal in several non-identical copies, that are individually equally-descriptive (with respect to a distortion metric such as the Mean Squared Error). The important aspect of the proposed idea is that subsets of the stored, non-identical, duplicates allow us to improve the quality of the recovered signal via a simple reconstruction procedure. Hence, the storage cost increase on the duplicates is exploited for significant quality improvement in the retrieved signals.

We design the framework for production of holographic representations employing binary compressed data. Since many compression processes are shift sensitive (e.g., due to block-based designs), we create holographic representations based on various shifts of the input signal.  Then, in the reconstruction stage the signal is approximated via averaging the available subset of properly back-shifted representations. The reconstruction quality improves as the  subset of available representations gets larger.

We further improve our idea by formulating the problem as a rate-distortion optimization, minimizing a Lagrangian cost including the total bit-cost of all the representations and two distortion penalties: one expresses the distortion averaged over all the $ m $-packet reconstructions (for a specific $ m > 1 $), and the second reflects the average distortion of individual packets. Then, we apply our general optimization approach for intricate compression problems (established in \cite{dar2018optimized,dar2018restoration,dar2018system,dar2018compression} for various settings). Specifically, using the alternating direction method of multipliers (ADMM) we develop an iterative process relying on repeated applications of standard compression techniques (that consider squared-error metrics but no holographic-representations aspects). Accordingly, our iterative approach decouples the holographic-related distortion terms from the actual compression stage, leading to holographic compressed representations compatible to an existing compression standard.

We present experimental results evaluating the proposed methodology for image compression in conjunction with the JPEG2000 standard. The results are analyzed using empirical quantities reflecting the holographic properties of similar usefulness of packets added to the reconstruction, as well as progressive refinement. 
Impressive PSNR gains are achieved by the proposed methods over the approach of exact duplications. For instance, we evaluate the case of four packets compatible with the JPEG2000 standard at a compression ratio of 1:50, and show that \textit{using four packets} the proposed optimization framework improves the PSNR of the reconstructed image by about 5 dB over the PSNR obtained with exact duplications.

\section{Problem Definition}
\label{sec:Problem Definition}

\subsection{Holographic Compression and Decompression}
\label{subsec:Holographic Compression and Decompression - A General Description}

\begin{figure*}[]
	\centering
	\includegraphics[width=0.9\textwidth]{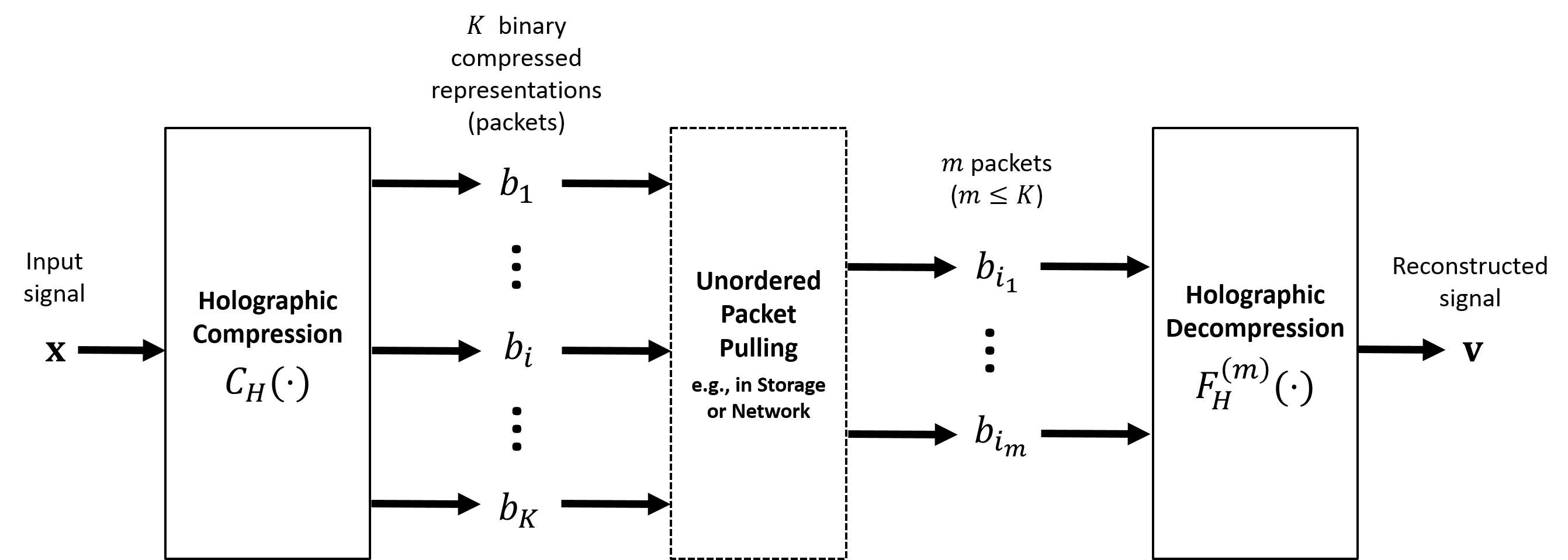}
	\caption{General description of holographic compression and decompression processes.} 
	\label{Fig:general_holographic_compression_decompression_process}
\end{figure*}

In this paper we propose a lossy compression framework with holographic representation properties (see Fig. \ref{Fig:general_holographic_compression_decompression_process}). 
Given a signal $ \vec{x} \in \mathbb{R}^N $, by definition, a holographic compression algorithm produces $ K $ binary representations (\textit{packets}) $ b_1 , ..., b_K \in \mathcal{B} $, where $ \mathcal{B} $ is a discrete set of binary compressed representations of possibly different lengths. The set of packets fulfill holographic properties either exactly or approximately (as will be described below). 
Accordingly, the holographic compression process can be described as a function $ C_H : \mathbb{R}^N  \rightarrow \mathcal{B}^K $, mapping the source signal domain, $\mathbb{R}^N $, to the $ K $-tuples from the domain $ \mathcal{B} $ of binary compressed representations.

By definition, the holographic decompression process can get any subset of $ m \in \left\lbrace 1,...,K \right\rbrace $ packets from the overall set of packets, a subset denoted here as $ \left\lbrace b_{i_1} , ..., b_{i_m} \right\rbrace \subset \left\lbrace  b_1 , ..., b_K \right\rbrace  $ where $ \left\lbrace {i_1}, ..., {i_m} \right\rbrace \subset \left\lbrace  1 , ..., K \right\rbrace$ are the indices of the packets taken from the range of integers from 1 to $ K $ without repetitions. For each $ m=1,...,K $ there is a holographic decompression function, $ F_H^{(m)} :  \mathcal{B}^m \rightarrow \mathbb{R}^N $, mapping the given subset of $ m $ packets into a reconstructed signal, namely, 
\begin{IEEEeqnarray}{rCl}
	\label{eq:holographic decompression - binary inputs}
	\vec{v} \triangleq F_H^{(m)} \left( b_{i_1} , ..., b_{i_m} \right)
\end{IEEEeqnarray}
where $ \vec{v} \in \mathbb{R}^N $.

We evaluate the fidelity of the reconstructed signal using the Mean Squared Error (MSE) criterion. Accordingly, the distortion of the reconstruction from the $ m $ packets corresponding to the indices $ i_1,...,i_m $ is formulated as 
\begin{IEEEeqnarray}{rCl}
	\label{eq:general description - MSE for reconstruction using m packets}
	\tilde{D}^{(m)} \left( \vec{x}; {i_1} , ..., {i_m} \right) \triangleq \frac{1}{N} \left\| { \vec{x}  - F_H^{(m)} \left( b_{i_1} , ..., b_{i_m} \right)  } \right\|_2^2 . 
\end{IEEEeqnarray}

In the sequel we will use the following notations. The sequence of integers from $ 1 $ to $ K $ is denoted as $ \integerSequence{K} \triangleq \nolinebreak \lbrace 1,...,K \rbrace $. For $ m \in \integerSequence{K} $, an $ m $-combination of the set $ \integerSequence{K} $ is a subset of $ m $ distinct numbers from $ \integerSequence{K} $. We denote the set of all $ m $-combinations of $ \integerSequence{K} $ as $ \integerSequence{K} \choose m $, where the latter contains $ K \choose m $ elements.

\subsection{The Ideal Holographic Properties in Deterministic Settings}
The desired holographic properties, in their idealistic forms, can be described as follows. 
\subsubsection{\textbf{Equivalent usefulness of individual packets}} 
Each of the individual packets,  $ \lbrace b_i \rbrace_{i=1}^K $, should enable the approximation of $ \vec{x} $ at the same level of MSE. More generally, given $ m\in \lbrace 1,..., K \rbrace $ packets, denoted as $ \lbrace {b}_{i_1} , ..., {b}_{i_m} \rbrace $, one can construct an estimate for $ \vec{x} $ using the function $F_H^{(m)} \left( b_{i_1} , ..., b_{i_m} \right) $  such that any subset of packets leads to a reconstruction that approximates $ \vec{x} $ at the same MSE level, i.e., this ideal property is formulated as 
\begin{IEEEeqnarray}{rCl}
	\label{eq:general description - holographic property of Equivalent usefulness of individual packets - definition}
	\tilde{D}^{(m)} \left( \vec{x}; {i_1} , ..., {i_m} \right) = \tilde{D}^{(m)} \left( \vec{x}; {l_1} , ..., {l_m} \right) 
\end{IEEEeqnarray}
for any $ \left( i_1, ..., i_m \right) $ and $ \left( l_1, ..., l_m \right) $ in  ${\integerSequence{K} \choose m }$.
\subsubsection{\textbf{Progressive refinement}}  
The approximation $F_H^{(m)} \left( b_{i_1} , ..., b_{i_m} \right) $ of $ \vec{x} $ using any $ m\in \lbrace 2,..., K \rbrace $ packets attains a lower MSE than the approximation $F_H^{(\bar {m})} \left( b_{i_1} , ..., b_{i_l} \right) $ constructed using any $ \bar {m} < \nolinebreak m $ packets. 
It is important to add the progressive refinement property to the equivalent usefulness concept or, otherwise, exact duplications of the input data would be a trivial solution to achieve equivalent usefulness of representations. 
The union of the above two properties can be formulated as follows: for $ m=1,...,K $,  
\begin{IEEEeqnarray}{rCl}
	\label{eq:general description - holographic property of progressive refinement - definition}
	\tilde{D}^{(m)} \left( \vec{x}; {i_1} , ..., {i_m} \right) =  \mathcal{E}_m  ~~~~ \forall \left( i_1, ..., i_m \right) \in {\integerSequence{K} \choose m } ~~~~~~~
\end{IEEEeqnarray}
where $ \mathcal{E}_{\bar {m}} > \mathcal{E}_{m} $ for any $ \bar {m} < m $ in $ \integerSequence{K} $.

\subsection{Feasible Holographic Properties in Deterministic Settings}

In general (in the deterministic settings), the \textit{ideal} holographic properties presented above cannot be precisely achieved. Hence, let us define a \textit{feasible} version of the holographic principles. 

First let us define the average MSE of the $ m $-packet reconstructions as 
\begin{IEEEeqnarray}{rCl}
	\label{eq:general description - feasible holographic properties - average MSE}
	&& \meanDistortion{m} \left( \vec{x}; b_{1} , ..., b_{K} \right) \triangleq 
	\\ \nonumber 
	&& \frac{1}{ {K \choose m } }  \sum_{ \left( i_1, ..., i_m \right) \in {\integerSequence{K} \choose m }  } \tilde{D}^{(m)} \left( \vec{x}; {i_1} , ..., {i_m} \right)
\end{IEEEeqnarray}
Furthermore, the empirical variance of the $ m $-packet reconstruction MSE is defined via 
\begin{IEEEeqnarray}{rCl}
	\label{eq:general description - feasible holographic properties - MSE variance}
	&& \varDistortion{m} \left( \vec{x}; b_{1} , ..., b_{K} \right) \triangleq \frac{1}{ {K \choose m } } \times 
	\\ \nonumber
	&& \sum_{ \substack{\left( i_1, ..., i_m \right) \\ \in {\integerSequence{K} \choose m } } } \left( \tilde{D}^{(m)} \left( \vec{x}; {i_1} , ..., {i_m} \right) - \meanDistortion{m} \left( \vec{x}; b_{1} , ..., b_{K} \right)  \right)^2
\end{IEEEeqnarray}

The definitions of average and variance of the reconstruction MSE allow us to formulate softened versions of the strict holographic properties defined in the former subsection. These practical features are 
\subsubsection{\textbf{$ \sigma $-Similar usefulness of individual packets}} 
Consider the task of reconstructions based on subsets of  $ m\in \lbrace 2,..., K \rbrace $ packets. A set of $ K $ packets, $ \lbrace b_i \rbrace_{i=1}^K $, will be considered to satisfy the property of $ \sigma $-similar usefulness of packets for $ m $-packet reconstructions, if it obeys 
\begin{IEEEeqnarray}{rCl}
	\label{eq:general description - feasible holographic properties - similar usefulness}
	\varDistortion{m} \left( \vec{x}; b_{1} , ..., b_{K} \right) \le \sigma^2 .
\end{IEEEeqnarray}
Namely, the variance of the reconstruction MSE, empirically considering all the $ m $-combinations of subsets, does not exceed the value $ \sigma^2 $.
Clearly, for $ \sigma = 0 $ the property defined here reduces to the strict equivalence of packet usefulness presented in (\ref{eq:general description - holographic property of Equivalent usefulness of individual packets - definition}).

\subsubsection{\textbf{Progressive refinement on average}}  
This property is implemented by a set of $ K $ packets where the approximations of $ \vec{x} $ using $ m\in \lbrace 2,..., K \rbrace $ packets yield a lower \textit{average} MSE than the approximations constructed using $ \bar {m} < \nolinebreak m $ packets. Namely, 
\begin{IEEEeqnarray}{rCl}
	\label{eq:general description - feasible holographic properties - progressive refinement in average}
	\meanDistortion{m} \left( \vec{x}; b_{1} , ..., b_{K} \right) =  {\mathcal{E}}_m  
\end{IEEEeqnarray}
where $ \mathcal{E}_{\bar {m}} > \mathcal{E}_{m} $ for any $ \bar {m} < m $ in $ \integerSequence{K} $.
It is again worth noting the significance of demanding progressive refinement (on average) in conjunction with the similar-usefulness concept, or else exact duplications of the input data would trivially provide equivalent usefulness of representations.

\section{Shift-Based Holographic Compression: A Baseline Approach}
\label{sec:Shift-Based Holographic Compression: A Baseline Approach}

\begin{figure*}[]
	\centering
	\includegraphics[width=0.95\textwidth]{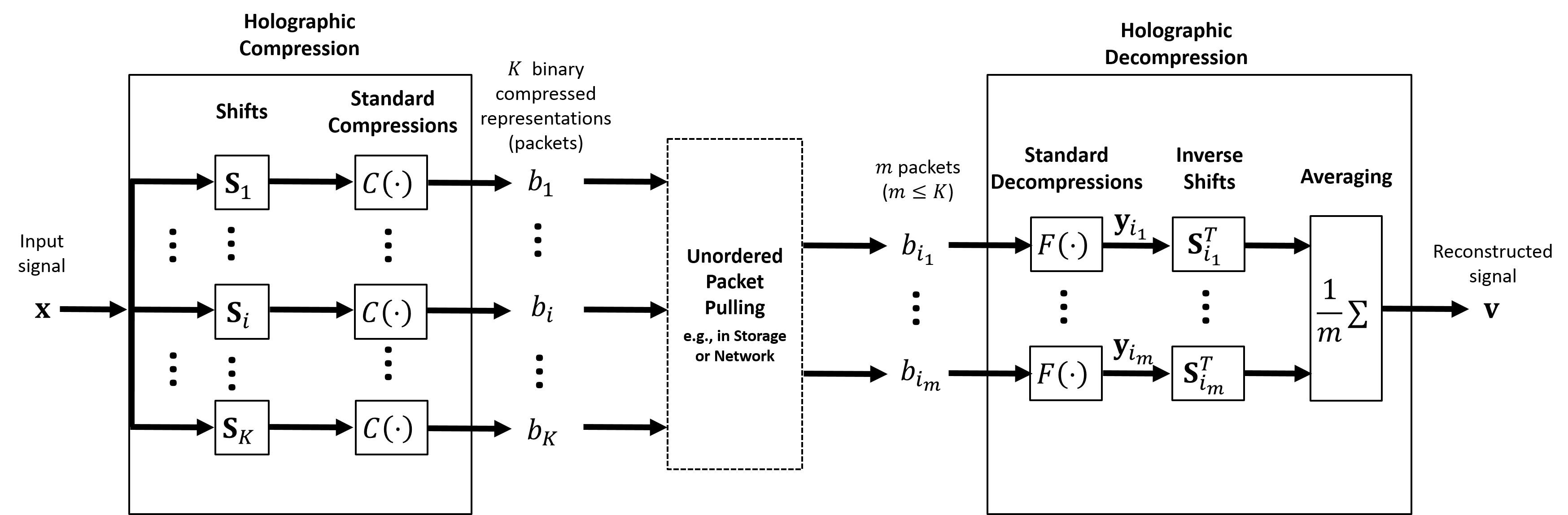}
	\caption{The baseline unoptimized process for holographic compression and decompression.} 
	\label{Fig:unoptimized_holographic_compression_decompression_process}
\end{figure*}

We next describe an elementary, yet effective, design for holographic compression. The simplicity of this baseline architecture stems from the utilization of shift operators in conjunction with standard compression methods that are inherently shift-sensitive. Specifically, the regular compression of the various shifts of the given signal will produce different compressed representations that are, in principle, of about the same usefulness for reconstruction. The progressive refinement ability is also immediate here due to the collection of different decompressed signals that, together, can provide a reconstruction with a lower distortion and reduced amount of compression artifacts.

For a start, let us formulate a process of regular (non holographic) lossy compression as a mapping $ C: \mathbb{R}^N \rightarrow \mathcal{B} $ from the $ N $-dimensional signal domain to a discrete set $ \mathcal{B} $ of binary compressed representations (of possibly different lengths) supported by the compression architecture. The compression of the signal $ \vec{w} \in \mathbb{R}^N $ provides the compressed binary data $\textit{b} = C \left( \vec{w} \right)$ that can be decompressed to form the signal $\vec{y} = F \left( \textit{b} \right)$, where $ F: \mathcal{B} \rightarrow \mathcal{S} $ represents the decompression mapping between the binary compressed representations in $ \mathcal{B} $ to the corresponding decompressed signals in the discrete set $ \mathcal{S} \subset \mathbb{R}^N $. Accordingly, we consider the pair of sets $ \mathcal{B} $ and $ \mathcal{S} $ as a description of a standard non-holographic compression architecture. 

Note that we intentionally associated the holographic compression design in Section \ref{subsec:Holographic Compression and Decompression - A General Description} with the standard compression definition given here, by referring to the same set $ \mathcal{B} $ of binary compressed representations. Indeed, this means that the holographic decompression process should start with individual standard decompression of the obtained packets, namely, 
\begin{IEEEeqnarray}{rCl}
	\label{eq:standard decompression of packets}
	\vec{y}_{j} = F\left( b_j \right)  \text{~~for~~} j=i_1, ...,i_m
\end{IEEEeqnarray}
where $ \vec{y}_{j} $ is the decompressed signal associated with the $ j^{th} $ packet. We will refer to $ \vec{y}_{i_1}, ..., \vec{y}_{i_m} $ as \textit{decompressed packets}.
Since the holographic decompression, associated with the function $ F_H^{(m)} $ defined in (\ref{eq:holographic decompression - binary inputs}), starts with standard decompression of the individual packets, we can define the relation 
\begin{IEEEeqnarray}{rCl}
	\label{eq:holographic decompression - relation between binary and signal inputs}
	G_H^{(m)} \left( \vec{y}_{i_1} , ..., \vec{y}_{i_m} \right) \triangleq F_H^{(m)} \left( b_{i_1} , ..., b_{i_m} \right)
\end{IEEEeqnarray}
i.e., $ G_H^{(m)}: \mathcal{S}^m \rightarrow \mathbb{R}^N $ is the holographic reconstruction function, receiving $ m $ decompressed packets and returning the decompressed signal $ \vec{v} \in \mathbb{R}^N $. For simplicity of notations, the developments in this paper mainly refer to the holographic decompression function $ G_H^{(m)}$ having inputs and outputs in the signal domain $ \mathbb{R}^N $.

Signal compression methods usually rely on various block-based vector quantization designs that inherently make them shift sensitive. Accordingly, we consider in this paper the creation of holographic compressed representations based on shift operators coupled with standard compression techniques. 
For this purpose we define the operator of a cyclic shift, to cyclically move components of an $ N $-length column vector in one place upward, via the $ N\times N $ matrix 
\begin{IEEEeqnarray}{rCl}
	\label{eq:cyclic shift - definition - single}
	\mtx{S} \triangleq \begin{bmatrix}
		0 & 1 & 0 & \cdots & 0 \\
		0 & 0 & 1 & \cdots & 0 \\
		\vdots  & \vdots  & \vdots & \ddots & \vdots  \\
		0 & 0 & 0 & \cdots & 1 \\
		1 & 0 & 0 & \cdots & 0
		\end{bmatrix}
\end{IEEEeqnarray}
and the corresponding inverse shift can be applied using $ \mtx{S}^T $ since $ \mtx{S}^T \mtx{S} = \mtx{I} $. The cyclic shift in an amount of $ l $ places is obtained via $ \mtx{S}^l $, which is the product of $ l $ basic matrices $ \mtx{S} $, and its inverse is accordingly defined as the transpose of  $ \mtx{S}^l $. 

As a baseline unoptimized design let us consider the following implementations of the holographic compression and decompression processes (see Fig. \ref{Fig:unoptimized_holographic_compression_decompression_process}). The holographic compression procedure $ C_H \left( \vec{x} \right) $ produces the $ K $ binary compressed representations via 
\begin{IEEEeqnarray}{rCl}
	\label{eq:unoptimized design - binary compressed representations}
	b_{i} = C\left( \mtx{S}_i \vec{x} \right)  \text{~~for~~} i=1, ...,K
\end{IEEEeqnarray}
where $ \mtx{S}_1, ...,\mtx{S}_K $ are $ K $ different cyclic shift operators in the forms of $ N \times N $ matrices. Accordingly, in this baseline architecture, the $ i ^{th} $ holographic compressed representation is formed by a standard compression of a (cyclically) shifted version of the input $ \vec{x} $ (where the amount of shift is defined by the matrix $ \mtx{S}_i $).
The holographic decompression based on a subset of $ m $ packets is defined as 
\begin{IEEEeqnarray}{rCl}
	\label{eq:unoptimized design - decompression from binary compressed representations}
	F_H^{(m)} \left( b_{i_1} , ..., b_{i_m} \right) = \frac{1}{m} \sum\limits_{j=1}^{m} \mtx{S}_{i_j}^T F\left( b_{i_j} \right)
\end{IEEEeqnarray}
or, alternatively, by describing the reconstruction given the decompressed packets as 
\begin{IEEEeqnarray}{rCl}
	\label{eq:unoptimized design - decompression from decompressed packets compressed representations}
	G_H^{(m)} \left( \vec{y}_{i_1} , ..., \vec{y}_{i_m} \right) = \frac{1}{m} \sum\limits_{j=1}^{m} \mtx{S}_{i_j}^T \vec{y}_{i_j} 
\end{IEEEeqnarray}

The MSE of the reconstruction from the $ m $ packets corresponding to the indices $ i_1,...,i_m $ is  
\begin{IEEEeqnarray}{rCl}
	\label{eq:MSE for reconstruction using m packets}
	D^{(m)} \left( \vec{x}; \vec{y}_{i_1} , ..., \vec{y}_{i_m} \right) \triangleq \frac{1}{N} \left\| { \vec{x}  - \frac{1}{m} \sum\limits_{ j=1}^{m} \mtx{S}_{i_j}^T \vec{y}_{i_j}  } \right\|_2^2 , 
\end{IEEEeqnarray}
where we use a simplified notation assuming that the indices of the packets (i.e., $ {i_1} , ..., {i_m} $) are available to the distortion function in order to associate the shift operators corresponding to the decompressed packets.

\section{An Optimization-Based Approach for Holographic Compression}
\label{sec:An Optimization-Based Approach for Holographic Compression}

Returning to the baseline implementation described in (\ref{eq:unoptimized design - binary compressed representations})-(\ref{eq:unoptimized design - decompression from decompressed packets compressed representations}) clearly shows that while the baseline design is a new and intriguing compression approach, it is not designed to optimize the output quality. The main goal of this section is to present an \textbf{optimized design for holographic compression based on the same, relatively simple, reconstruction procedures in (\ref{eq:unoptimized design - decompression from binary compressed representations})-(\ref{eq:unoptimized design - decompression from decompressed packets compressed representations}), while replacing the encoding process of (\ref{eq:unoptimized design - binary compressed representations}) by our optimization-induced procedure.}

We now turn to define the holographic compression problem in the form of a rate-distortion optimization, posed for improving the average quality of $ m $-packet reconstructions for a specific $ m \in \lbrace 2,...,K\rbrace $. 
Our initial problem formulation is inspired by the rate-distortion Lagrangian optimization that is commonly used in the state-of-the-art image and video compression methods (see, for examples, \cite{shoham1988efficient,ortega1998rate,sullivan1998rate,sullivan2012overview}). 
Here we formulate the task as the minimization of an extended rate-distortion Lagrangian cost, including three main terms: the total compression bit-cost of the packets, the average MSE of $ m $-packet reconstructions (defined for a particular $ m \in \lbrace 2,...,K\rbrace $), and the average MSE of reconstructions from individual packets. This optimization is formulated as 
\begin{IEEEeqnarray}{rCl}
	\label{eq: holographic compression - relaxed problem - unconstrained Lagrangian}
	&& \lbrace \hat{\vec{y}}_i \rbrace_{i=1}^{K}  = \underset{  { \lbrace \vec{y}_i \rbrace_{i=1}^{K} \in \mathcal{S} } }{\text{argmin}}  ~ { \sum\limits_{i=1 }^{K} R\left( \vec{y}_i \right) } \\ \nonumber
	&& \qquad\qquad + \mu \frac{1}{ {K \choose m } }  \sum_{ \left( i_1, ..., i_m \right) \in {\integerSequence{K} \choose m }  } D^{(m)} \left( \vec{x}; \vec{y}_{i_1} , ..., \vec{y}_{i_m} \right) \\ \nonumber
	&& \qquad\qquad +  \lambda \frac{1}{K} \sum\limits_{i=1}^{K}  D^{(1)} \left( \vec{x}; \vec{y}_i \right)
\end{IEEEeqnarray}
where $ \mu $ and $ \lambda $ are Lagrange multipliers corresponding to some trade-off among the bit-cost and the distortion quantities. 
It is important to note that the reduction in the average MSE of $ m $-packet reconstructions usually leads to increase in the average MSE of individual-packet reconstructions. Therefore, in our experiments (see Section \ref{sec:Experimental Results}) we will set the values of $ \mu $ and $ \lambda $ such that the average MSE of $ m $-packet reconstructions will be the desired distortion value to minimize, and the inclusion of the average MSE of individual-packet reconstructions is for regularization purposes, namely, to limit the degradation introduced to single-packet representations. This aspect of the optimization is clearly exhibited in the empirical demonstrations provided in Section \ref{sec:Experimental Results}. 

We suggest to address the optimization in (\ref{eq: holographic compression - relaxed problem - unconstrained Lagrangian}) using the alternating direction method of multipliers (ADMM) approach \cite{boyd2011distributed}. For a start, we apply variable splitting on the optimization in (\ref{eq: holographic compression - relaxed problem - unconstrained Lagrangian}), translating the problem to 
\begin{IEEEeqnarray}{rCl}
	\label{eq: holographic compression - relaxed problem - variable splitting}
	&& \left(  \lbrace \hat{\vec{y}}_i \rbrace_{i=1}^{K} , \lbrace \hat{\vec{z}}_i \rbrace_{i=1}^{K} \right) = \underset{ \substack{ { \lbrace \vec{y}_i \rbrace_{i=1}^{K} \in \mathcal{S} ,}\\{ \lbrace \vec{z}_i \rbrace_{i=1}^{K} \in \mathbb{R}^N }} }{\text{argmin}}   { \sum\limits_{i=1 }^{K} R\left( \vec{y}_i \right) } \nonumber \\ \nonumber
	&& \qquad\qquad + \mu \frac{1}{ {K \choose m } }  \sum_{ \left( i_1, ..., i_m \right) \in {\integerSequence{K} \choose m }  } D^{(m)} \left( \vec{x}; \vec{z}_{i_1} , ..., \vec{z}_{i_m} \right) \\ \nonumber
	&& \qquad\qquad +  \lambda \frac{1}{K} \sum\limits_{i=1}^{K}  D^{(1)} \left( \vec{x}; \vec{z}_i \right)
	\\
	&& \text{subject to} ~~  \vec{z}_i = \vec{y}_i ~~ \forall ~ i \in  \integerSequence{K} 
\end{IEEEeqnarray}
where $ \vec{z}_1, ... , \vec{z}_K $ are auxiliary variables, which are not directly restricted to the discrete set $ \mathcal{S} $.
Then, the augmented Lagrangian and the method of multipliers \cite{boyd2011distributed} provide an iterative form of the problem where its $ t^{th} $ iteration is formulated as 
\begin{IEEEeqnarray}{rCl}
	\label{eq: holographic compression - relaxed problem - augmented Lagrangian}
	&& \left(  \lbrace \hat{\vec{y}}_i^{[t]} \rbrace_{i=1}^{K} , \lbrace \hat{\vec{z}}_i^{[t]} \rbrace_{i=1}^{K} \right) = \underset{ \substack{ { \lbrace \vec{y}_i \rbrace_{i=1}^{K} \in \mathcal{S} ,}\\{ \lbrace \vec{z}_i \rbrace_{i=1}^{K} \in \mathbb{R}^N }} }{\text{argmin}}  { \sum\limits_{i=1 }^{K} R\left( \vec{y}_i \right) } + \nonumber \\ \nonumber
&& \qquad\qquad + \mu \frac{1}{ {K \choose m } }  \sum_{ \left( i_1, ..., i_m \right) \in {\integerSequence{K} \choose m }  } D^{(m)} \left( \vec{x}; \vec{z}_{i_1} , ..., \vec{z}_{i_m} \right) \\ \nonumber
&& \qquad\qquad +  \lambda \frac{1}{K} \sum\limits_{i=1}^{K}  D^{(1)} \left( \vec{x}; \vec{z}_i \right)
\\ \nonumber
	&& \qquad\qquad + \beta \sum\limits_{i=1}^{K} \left\| { \vec{y}_i  - \vec{z}_i + \vec{u}_i^{[t]} }\right\|_2^2
	\\
	&& \vec{u}_i^{[t+1]} = \vec{u}_i^{[t]} + \left( \hat{\vec{y}}_i^{[t]} - \hat{\vec{z}}_i^{[t]} \right) ~~\forall ~ i \in  \integerSequence{K}
\end{IEEEeqnarray}
where $ \beta $ is a parameter originating in the augmented Lagrangian, and $ \vec{u}_1^{[t]}, ..., \vec{u}_K^{[t]} $ are scaled dual variables. We denote correspondence to specific iterations using superscript square-brackets, whereas other types of superscripts (e.g., including round brackets) correspond to former definitions given above.

Addressing the optimization in (\ref{eq: holographic compression - relaxed problem - augmented Lagrangian}) using one iteration of alternating minimization establishes the following ADMM form of the problem, where its $ t^{th} $ iteration is 
\begin{IEEEeqnarray}{rCl}
	\label{eq: holographic compression - relaxed problem - ADMM - compression}
	&& \hat{\vec{y}}_i^{[t]} = \underset{ \vec{y}_i \in \mathcal{S} }{\text{argmin}} ~ { R\left( \vec{y}_i \right) } + \beta \left\| { \vec{y}_i  - \tilde{\vec{z}}_i^{[t]} }\right\|_2^2 ~~\forall ~ i \in  \integerSequence{K}
	\\
	\label{eq: holographic compression - relaxed problem - ADMM - L2 terms}
	&& \hat{\vec{z}}_i^{[t]}  = \underset{ \vec{z}_i }{\text{argmin}}  ~  \frac{\mu}{ {K \choose m } }  \times \\ \nonumber 
	&& \sum_{ \left( i_1, ..., i_m \right) \in {\mathcal{I}_i^{(m)}}} D^{(m)} \left( \vec{x}; \lbrace \hat{\vec{z}}_{i_j}^{[t]} \rbrace_{ \substack{ {i_j < i} \\ {j\in\integerSequence{m} } } } , \vec{z}_{i} , \lbrace \hat{\vec{z}}_{i_j}^{[t-1]} \rbrace_{ \substack{ {i_j > i} \\ {j\in\integerSequence{m} } } } \right) \nonumber \\
	&& +  \frac{\lambda}{K} D^{(1)} \left( \vec{x}; \vec{z}_i \right) +  \beta \left\| {  \vec{z}_i - \tilde{\vec{y}}_i^{[t]} }\right\|_2^2  ~~\forall ~ i \in  \integerSequence{K} \nonumber
	\\ \label{eq: holographic compression - relaxed problem - ADMM - update dual variables}
	&& \vec{u}_i^{[t+1]} = \vec{u}_i^{[t]} + \left( \hat{\vec{y}}_i^{[t]} - \hat{\vec{z}}_i^{[t]} \right) ~~\forall ~ i \in  \integerSequence{K}
\end{IEEEeqnarray}
where $ \tilde{\vec{z}}_i^{[t]} \triangleq \hat{\vec{z}}_i^{[t-1]} - \vec{u}_i^{[t]} $ and $ \tilde{\vec{y}}_i^{[t]} \triangleq \hat{\vec{y}}_i^{[t]} + \vec{u}_i^{[t]} $. Moreover, the optimization of $ \vec{z}_i $ in (\ref{eq: holographic compression - relaxed problem - ADMM - L2 terms}) considers the average reconstruction MSE corresponding to all the $ m $-combinations of packets including the $ i^{th} $ packet  -- the set of these $ m $-combinations is denoted as $ \mathcal{I}_i^{(m)} $. Note also that the size of this set is $ \lvert \mathcal{I}_i^{(m)} \rvert = {K-1 \choose m-1} $.

The optimizations in (\ref{eq: holographic compression - relaxed problem - ADMM - compression}) are standard rate-distortion optimizations with respect to a squared error metric, considering the individual compression of $ \tilde{\vec{z}}_i^{[t]} $ for each $ i=1,...,K $. Therefore, we suggest to replace the optimizations in (\ref{eq: holographic compression - relaxed problem - ADMM - compression}) with applications of standard compression and decompression operated based on a parameter $ \theta \left( \beta \right) $ determining the bit-rate (see stage 8 of Algorithm \ref{Algorithm:Proposed Method}). For example, the experiments presented in Section \ref{sec:Experimental Results} leverage the JPEG2000 compression technique, applied using a compression-ratio parameter. Interestingly, in our experiments we find it sufficient to set $ \theta \left( \beta \right) $ to a constant value (heuristically determined based on the $ \beta $ value) and kept fixed throughout the iterations (i.e., $ \theta \left( \beta \right) $ is considered to be independent of $ t $).

The second optimization stage, Eq. (\ref{eq: holographic compression - relaxed problem - ADMM - L2 terms}), can be analytically solved with respect to the explicit expressions provided in (\ref{eq:MSE for reconstruction using m packets}) for the distortion measures, showing that 
\begin{IEEEeqnarray}{rCl}
	\label{eq: holographic compression - relaxed problem - ADMM - L2 terms - analytic solution}
	&& \hat{\vec{z}}_i^{[t]}  = \frac{ N \beta \tilde{\vec{y}}_i^{[t]} + \frac{\lambda}{K} \mtx{S}_i \vec{x} + \frac{\mu}{ m^2 \cdot {K \choose m } } \mtx{S}_i \vec{w}_i^{(m)}   }{N \beta + \frac{\lambda}{K} + \frac{\mu}{ m^2 \cdot {K \choose m } } \cdot {\lvert \mathcal{I}_i^{(m)} \rvert} } 
\end{IEEEeqnarray}
where 
\begin{IEEEeqnarray}{rCl}
	\label{eq: holographic compression - relaxed problem - ADMM - L2 terms - analytic solution - auxiliary expression}
&& \vec{w}_i^{(m)} \triangleq  
\\ \nonumber 
&& \sum_{ \left( i_1, ..., i_m \right) \in {\mathcal{I}_i^{(m)}}} \left( m \vec{x}  -  \sum\limits_{{ \substack{ {i_j < i} \\ {j\in\integerSequence{m} } } }} \mtx{S}_{i_j}^T  \hat{\vec{z}}_{i_j}^{[t]}  - \sum\limits_{{ \substack{ {i_j > i} \\ {j\in\integerSequence{m} } } }} \mtx{S}_{i_j}^T  \hat{\vec{z}}_{i_j}^{[t-1]} \right) . 
\end{IEEEeqnarray}
The expression in (\ref{eq: holographic compression - relaxed problem - ADMM - L2 terms - analytic solution}) exhibits $ \hat{\vec{z}}_i^{[t]} $ as a linear combination of the corresponding decompressed packet $ \tilde{\vec{y}}_i^{[t]} $, the shifted input signal $ \vec{x} $, and the shifted residual between $ \vec{x} $ and its $ m $-packet approximations excluding the $ i^{th} $ packet. 
Note that in the case of $ m=K $ (namely, optimizing the reconstruction using all the packets), the expression in (\ref{eq: holographic compression - relaxed problem - ADMM - L2 terms - analytic solution - auxiliary expression}) is somewhat simplified to 
\begin{IEEEeqnarray}{rCl}
	\label{eq: holographic compression - relaxed problem - ADMM - L2 terms - analytic solution - optimize reconstruction from all}
	\vec{w}_i^{(K)} =  K \vec{x}  -  \sum\limits_{j=1}^{i-1} \mtx{S}_j^T  \hat{\vec{z}}_j^{[t]}  - \sum\limits_{j=i+1}^{K} \mtx{S}_j^T  \hat{\vec{z}}_j^{[t-1]} . 
\end{IEEEeqnarray}

The method developed in this section is summarized in Algorithm \ref{Algorithm:Proposed Method}, where the processing of packets in each iteration is done sequentially. This reordering of computations is allowed due to the formation of dependencies obtained in Eq. (\ref{eq: holographic compression - relaxed problem - ADMM - compression})-(\ref{eq: holographic compression - relaxed problem - ADMM - update dual variables}).

\begin{algorithm}
	\caption{Holographic Compression Optimized for $ m $-Packet Reconstructions}
	\label{Algorithm:Proposed Method}
	\begin{algorithmic}[1]
		\State Inputs: $ \vec{x} $, $ {\beta} $, $ \mu $, $ \lambda $, $ m $, $ K $.
		\State  Initialize $t = 0$.
		\State  Initialize (for $ i=1,...,K $) $  {\hat{\vec{z}}}_i^{(0)} = \mtx{S}_i \vec{x} $ and $\vec{u}_i^{(1)} = \vec{0}$. 
		\Repeat 
		
		\State $ t \gets t + 1$

		\For{$ i=1,...,K $}
		\State $ \tilde{ \vec{z}}_i^{[t]} = \hat{\vec{z}}_i^{[t-1]} - \vec{u}_i^{[t]} $
		
		\State $ {\textit{b}}_i^{[t]} = StandardCompress \left( \tilde{ \vec{z}}_i^{[t]}, {\theta\left(\beta\right)} \right) $
		\State $ \hat{\vec{y}}_i^{[t]} = StandardDecompress \left( {\textit{b}}_i^{[t]} \right) $

		\State $ \tilde{ \vec{y}}_i^{[t]} = \hat{\vec{y}}_i^{[t]} + \vec{u}_i^{[t]} $
		\State $\hat{\vec{z}}_i^{[t]}  = \frac{ N \beta \tilde{\vec{y}}_i^{[t]} +  \frac{\lambda}{K} \mtx{S}_i \vec{x} + \frac{\mu}{ m^2 \cdot {K \choose m } } \mtx{S}_i \vec{w}_i^{(m)}   }{N \beta + \frac{\lambda}{K} + \frac{\mu}{ m^2 \cdot {K \choose m } } \cdot {\lvert \mathcal{I}_i^{(m)} \rvert} }$ 
		\Statex \qquad\qquad where $ \vec{w}_i^{(m)} $ is defined in (\ref{eq: holographic compression - relaxed problem - ADMM - L2 terms - analytic solution - auxiliary expression}).
		
		\State $\vec{u}_i^{[t+1]} = \vec{u}_i^{[t]} + \left( \hat{ \vec{y}}_i^{[t]} - \hat{\vec{z}}_i^{[t]} \right)$
		\EndFor
		
		\Until{stopping criterion is satisfied}
		\State Output: The binary compressed packets $ {\textit{b}}_1^{[t]},..., {\textit{b}}_K^{[t]} $.
	\end{algorithmic}
\end{algorithm}

\section{Experimental Results}
\label{sec:Experimental Results}

In this section we present experimental results for the implementation of the proposed method for holographic compressed representations of images in conjunction with the JPEG2000 compression technique (available in Matlab). 
In the presented evaluation we consider several settings for the storage of a given image using four copies (that are not necessarily identical) or packets. Each packet/copy is a compressed image in a binary form obtained from the JPEG2000 compression method operated at the same compression ratio. Therefore, all the individual copies and packets are of about the same bit-rate, allowing to evaluate reconstruction quality as the function of the number of packets/copies utilized. The four approaches examined here are: 
\begin{itemize}
	\item \textbf{\textit{Exact duplication}} where all the stored copies are exactly the same binary data, obtained from the JPEG2000 compression of the given image.
	\item \textbf{\textit{The baseline (unoptimized) design}}, as presented in Section \ref{sec:Shift-Based Holographic Compression: A Baseline Approach}, relying on JPEG2000 compression of different shifts of the input image. 
	\item \textbf{\textit{The shift-based holographic compression approach optimized for 2-packet reconstructions}}, as developed in Section \ref{sec:An Optimization-Based Approach for Holographic Compression} for optimizing the quality of $ m $-packet reconstructions. This design also relies on the JPEG2000 compression standard. The parameters for this mode are $ \mu = 25\cdot K \cdot {K \choose m } $, $ \beta = \frac{90}{N} $, $ \lambda = 5\cdot K^2  $, and a run of 35 iterations.
	\item \textbf{\textit{The shift-based holographic compression approach optimized for $ K $-packet reconstructions}}, namely, the case of optimizing the reconstruction using all the packets. The parameters for this mode are $ \mu = 125\cdot K \cdot {K \choose m } $, $ \beta = \frac{50}{N} $, $ \lambda = 2.5 \cdot K^2  $, and a run of 35 iterations.
\end{itemize}

\begin{figure}[]
	\centering
	\includegraphics[width=0.40\textwidth]{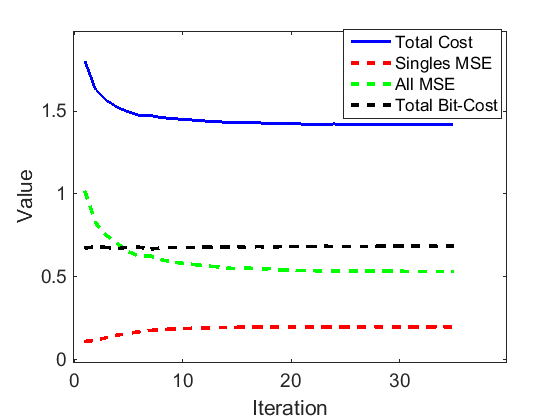}
	\caption{The evolution of the optimization cost and its components throughout the proposed iterative optimization. The demonstration here is for the Cameraman image and the optimization of 4-packet reconstruction composed of JPEG2000 packets having compression ratio of 1:50. The presented values of the cost-components include the multiplication by the respective parameters. } 
	\label{Fig:cameraman_jpeg2000_overall4packets_cost_evolution}
\end{figure}

The first evaluation is based on JPEG2000 compression at a compression ratio of 1:50 that in practice  creates packets at bit-rates of 0.160 bits per pixel (bpp), with the addition of some overhead bit-rate due to syntax (note that the overhead bit-rate is smaller for larger images). 
The baseline and the two optimized modes produce their four holographic packets based on the following offsets of the upper-left coordinate of the image by $ (0,0), (3,0), (0,3), (3,3)$ pixels (namely, in practice, the shifts are not cyclic and implemented by appending a suitable number of duplicated rows and columns at the upper and left sides of the image, respectively). 
The evolution of the optimization cost (formulated in Eq. (\ref{eq: holographic compression - relaxed problem - unconstrained Lagrangian})) and its components is demonstrated in Fig. \ref{Fig:cameraman_jpeg2000_overall4packets_cost_evolution}, showing the reduction in the optimization cost (the blue curve) and a convergence behavior.

\begin{figure*}[]
	\centering
	{\subfloat[1-packet reconstruct. using packet \#1 (23.04 dB)]{\label{fig:cameraman_jpeg2000_overall4packets_1packet_approx__packet_num1__23_04dB__0_170bpp}\includegraphics[width=0.23\textwidth]{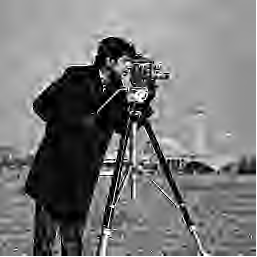}}}
	{\subfloat[1-packet reconstruct. using packet \#2 (23.02)]{\label{fig:cameraman_jpeg2000_overall4packets_1packet_approx__packet_num2__23_02dB__0_172bpp}\includegraphics[width=0.23\textwidth]{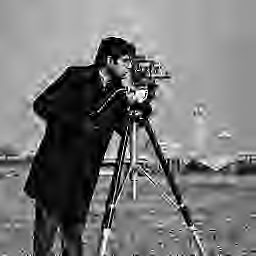}}}
	{\subfloat[1-packet reconstruct. using packet \#3 (22.96 dB)]{\label{fig:cameraman_jpeg2000_overall4packets_1packet_approx__packet_num3__22_96dB__0_172bpp}\includegraphics[width=0.23\textwidth]{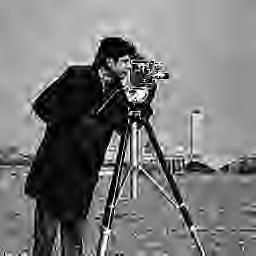}}}
	{\subfloat[1-packet reconstruct. using packet \#4  (23.01 dB)]{\label{fig:cameraman_jpeg2000_overall4packets_1packet_approx__packet_num4__23_01dB__0_172bpp}\includegraphics[width=0.23\textwidth]{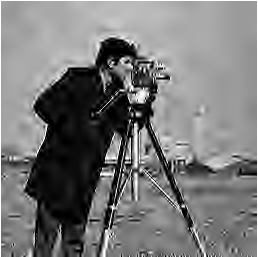}}}
\\
	{\subfloat[2-packet reconstruct.  (25.98 dB)]{\label{fig:cameraman_jpeg2000_overall4packets_2packet_approx__25_98dB__0_342bpp}\includegraphics[width=0.23\textwidth]{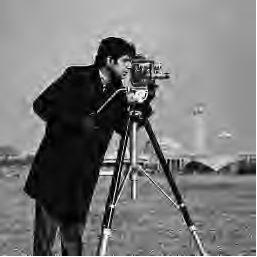}}}
	{\subfloat[3-packet reconstruct. (28.20)]{\label{fig:cameraman_jpeg2000_overall4packets_3packet_approx__28_20dB__0_514bpp}\includegraphics[width=0.23\textwidth]{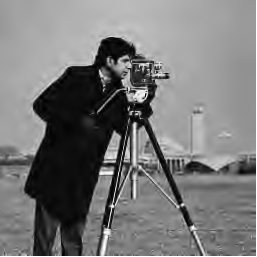}}}
	{\subfloat[4-packet reconstruct.  (29.73 dB)]{\label{fig:cameraman_jpeg2000_overall4packets_4packet_approx__29_73dB__0_687bpp}\includegraphics[width=0.23\textwidth]{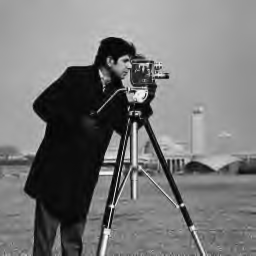}}}	
	\caption{Examples for $ m $-packet reconstructions of the 'Cameraman' image using multiple packets from the set of 4 holographic representations.
		Demonstration of $ m $-packet reconstructions obtained from a set of 4 holographic packets optimized by the proposed framework for a 4-packet reconstruction.  The utilized compression is JPEG2000 at a compression ratio of 1:50.
		(a)-(d) the 1-packet reconstructions using each of the individual packets. (e)-(g) examples for the $ m $-packet reconstructions for $ m=2,3,4 $.} 
	\label{Fig:Experiments - cameraman - optimized for 4 holographic representations}
\end{figure*}

In Fig. \ref{Fig:Experiments - cameraman - optimized for 4 holographic representations} we demonstrate the reconstructions obtained using the proposed holographic compression method optimized for 4-Packet reconstructions. First, in Fig. \ref{fig:cameraman_jpeg2000_overall4packets_1packet_approx__packet_num1__23_04dB__0_170bpp}-\ref{fig:cameraman_jpeg2000_overall4packets_1packet_approx__packet_num4__23_01dB__0_172bpp}, we present the reconstructions retrieved from each of the single packets alone: while the PSNR values are relatively similar, the approximations are clearly distinct and each of them suffers differently from compression artifacts. This observation explains the benefits from jointly using several packets for reconstruction. Then, in Fig. \ref{fig:cameraman_jpeg2000_overall4packets_2packet_approx__25_98dB__0_342bpp}-\ref{fig:cameraman_jpeg2000_overall4packets_4packet_approx__29_73dB__0_687bpp}, several examples for approximations using an increasing number of packets show the significance of the obtained improvements in PSNR and visual quality. 

\begin{figure*}[]
	\centering
	{\subfloat[Cameraman ]{\label{fig:cameraman_PSNR_curves}\includegraphics[width=0.24\textwidth]{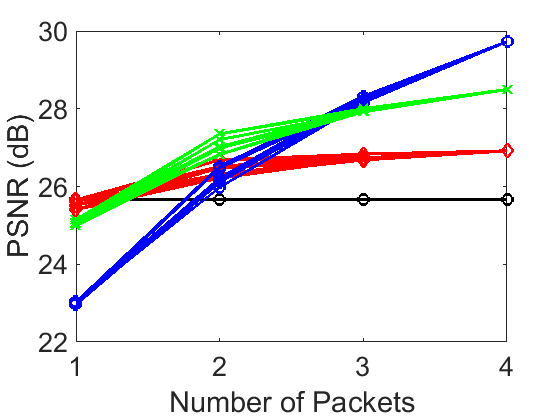}}}~~
	{\subfloat[House]{\label{fig:house_PSNR_curves}\includegraphics[width=0.24\textwidth]{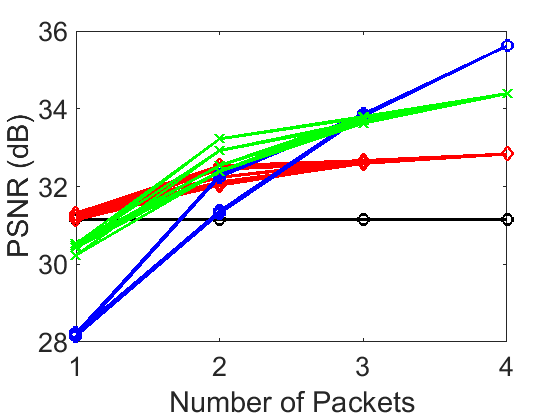}}}~~
	{\subfloat[Lena]{\label{fig:Lena512_PSNR_curves}\includegraphics[width=0.24\textwidth]{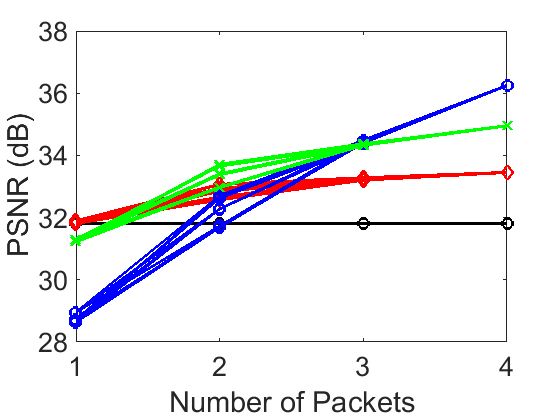}}}~~
	{\subfloat[Barbara]{\label{fig:Barbara_PSNR_curves}\includegraphics[width=0.24\textwidth]{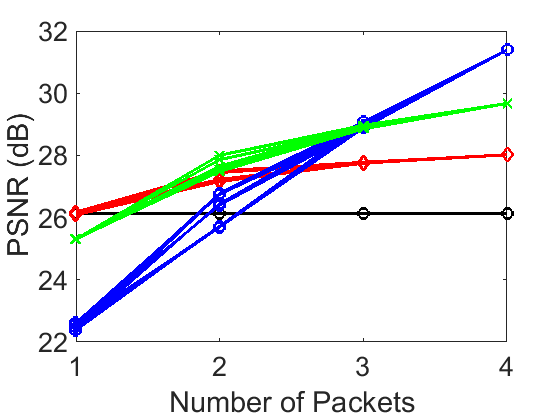}}}
	\caption{PSNR versus the number of packets used for the reconstructions. The complete set contains 4 packets, each obtained from JPEG2000 compression at 1:50 compression ratio. The black, red, green and blue curves respectively represent the methods of exact duplications, baseline (unoptimized), optimized for reconstruction from pairs of packets, and optimized for reconstruction from 4 packets. } 
	\label{Fig:Experiments - 4 holographic representations -  PSNR-number of packets curves}
\end{figure*}

Figure \ref{Fig:Experiments - 4 holographic representations -  PSNR-number of packets curves} allows to compare the examined methods through their corresponding curves of PSNR versus number of packets utilized for reconstruction. The results provided here are for the Cameraman (256$ \times $ 256 pixels), House (256$ \times $ 256 pixels), Lena (512$ \times $ 512 pixels) and Barbara (512$ \times $ 512 pixels) grayscale images. Each of the four examined methods is associated in Fig. \ref{Fig:Experiments - 4 holographic representations -  PSNR-number of packets curves} with a group of curves having the same color (which is method specific). The curves corresponding to a particular method differ by the order of appending packets for the reconstruction (and, therefore, the number of curves corresponding to each method is $ K! = 4! = 24 $). 
A good implementation of the holographic property of \text{similar usefulness} (see Section \ref{sec:Problem Definition}) means here that the diversity in PSNR values using the various combinations of $ m $ packets should be relatively small -- we quantify this in Table \ref{table:Experiments - blur - Average PSNR and Bit-Rate comparison} using the standard deviation of the PSNR obtained using the various subsets of $ m $ packets. The second important property is the \textit{progressive refinement} (see Section \ref{sec:Problem Definition}) that can be observed in the PSNR curves of all the shift-based holographic compression methods (see Fig. \ref{Fig:Experiments - 4 holographic representations -  PSNR-number of packets curves}), and is completely absent in the exact duplication approach. It is also evident that our optimization framework improves the average PSNR of the $ m $-packet reconstructions for the specific $ m $ set to be optimized (see Table \ref{table:Experiments - blur - Average PSNR and Bit-Rate comparison} and Fig. \ref{Fig:Experiments - 4 holographic representations -  PSNR-number of packets curves}). For instance, our optimization for $ 4 $-packet reconstructions achieved a PSNR gain of about 5 dB over the method of exact duplications, and a PSNR improvement around 3 dB over the baseline (unoptimized) shift-based approach. 

The presented comparison also demonstrates the fundamental, intuitive, trade-off in the average quality of $ m $-packet reconstructions among the various subset sizes $ m $. For example, the significant increase in the 4-packet reconstruction quality is at the expense of the qualities of the 1-packet reconstructions. Nonetheless, the optimizations for reconstructions using 4 or 2 packets indirectly led to significant improvement in the average quality of the 3-packet reconstructions in addition to the explicit optimization goal.

\begin{figure*}[]
	\centering
	{\subfloat[Cameraman ]{\label{fig:cameraman_PSNR_curves__compression_ratio_25}\includegraphics[width=0.24\textwidth]{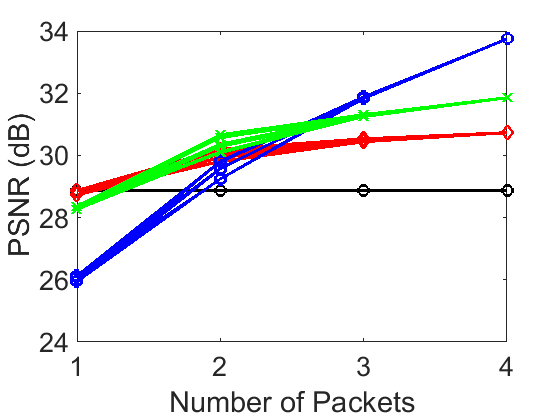}}}~~
	{\subfloat[House]{\label{fig:house_PSNR_curves__compression_ratio_25}\includegraphics[width=0.24\textwidth]{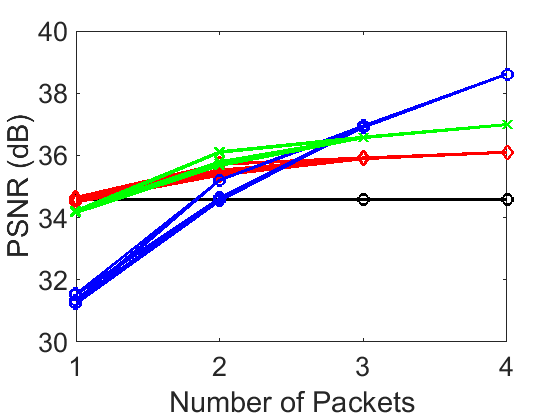}}}~~
	{\subfloat[Lena]{\label{fig:Lena512_PSNR_curves__compression_ratio_25}\includegraphics[width=0.24\textwidth]{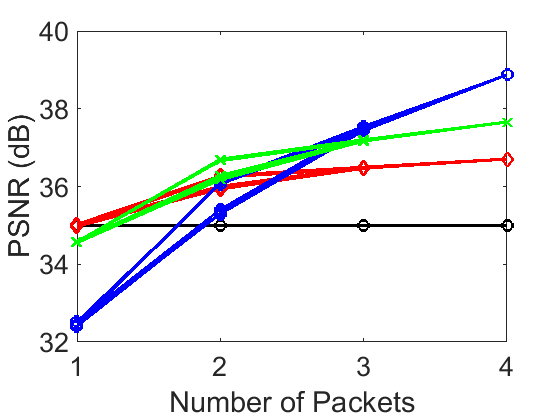}}}~~
	{\subfloat[Barbara]{\label{fig:Barbara_PSNR_curves__compression_ratio_25}\includegraphics[width=0.24\textwidth]{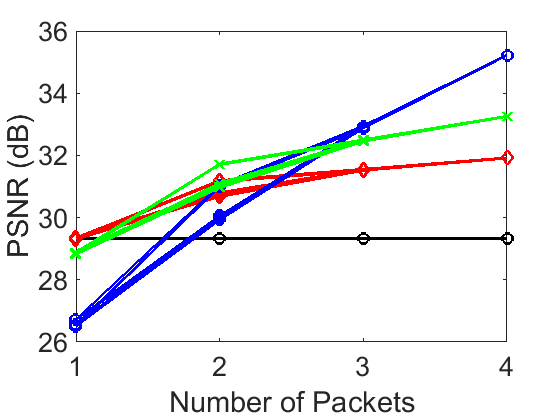}}}
	\caption{PSNR versus the number of packets used for the reconstructions. The complete set contains 4 packets, each obtained from JPEG2000 compression at 1:25 compression ratio. The black, red, green and blue curves respectively represent the methods of exact duplications, baseline (unoptimized), optimized for reconstruction from pairs of packets, and optimized for reconstruction from 4 packets.} 
	\label{Fig:Experiments - 4 holographic representations -  PSNR-number of packets curves - Compression Ratio 25 }
\end{figure*}

We repeat the experiment but for JPEG2000 compression at a ratio of 1:25, namely, a higher bit-rate of approximately 0.320 bits per pixel. The formulas for setting the parameters are as in the first setting described above, except for the $ \beta $ parameter, set in the 2-packet optimization mode to $ \frac{65}{N} $, and in the 2-packet optimization mode to $ \frac{120}{N} $.  The results are presented in Table \ref{table:Experiments - blur - Average PSNR and Bit-Rate comparison - Compression Ratio 25} and Fig. \ref{Fig:Experiments - 4 holographic representations -  PSNR-number of packets curves - Compression Ratio 25 }. Evidently, our framework consistently provides improved qualities of the reconstructions specified in the optimization task. 

\begin{figure*}[]
	\centering
	{\subfloat[Cameraman ]{\label{fig:cameraman_PSNR_curves__9packets_compression_ratio_50}\includegraphics[width=0.24\textwidth]{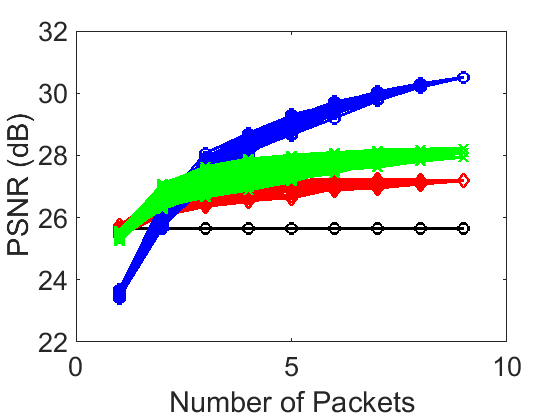}}}~~
	{\subfloat[House]{\label{fig:house_PSNR_curves__9packets_compression_ratio_50}\includegraphics[width=0.24\textwidth]{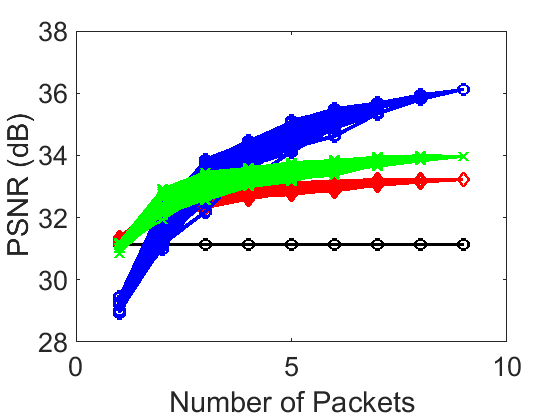}}}~~
	{\subfloat[Lena]{\label{fig:Lena512_PSNR_curves__9packets_compression_ratio_50}\includegraphics[width=0.24\textwidth]{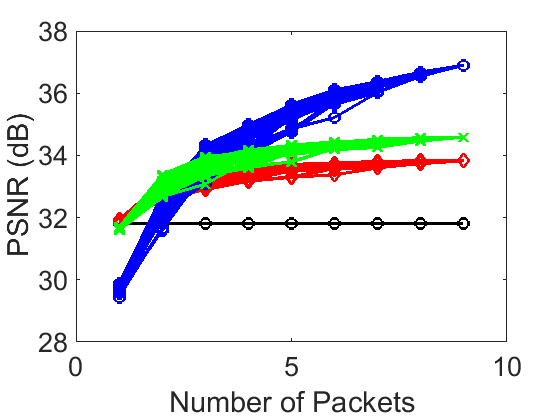}}}~~
	{\subfloat[Barbara]{\label{fig:Barbara_PSNR_curves__9packets_compression_ratio_50}\includegraphics[width=0.24\textwidth]{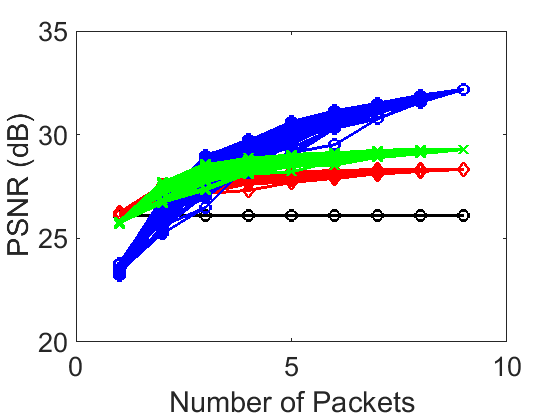}}}
	\caption{PSNR versus the number of packets used for the reconstructions. The complete set contains 9 packets, each obtained from JPEG2000 compression at 1:50 compression ratio. The black, red, green and blue curves respectively represent the methods of exact duplications, baseline (unoptimized), optimized for reconstruction from pairs of packets, and optimized for reconstruction from 9 packets.} 
	\label{Fig:Experiments - 9 holographic representations -  PSNR-number of packets curves}
\end{figure*}

\begin{figure*}[]
	\centering
	{\subfloat[1-packet reconstruction (23.81 dB)]{\label{fig:barbara_jpeg2000_overall9packets_1packet_approx__23_81dB__0_153bpp}\includegraphics[width=0.31\textwidth]{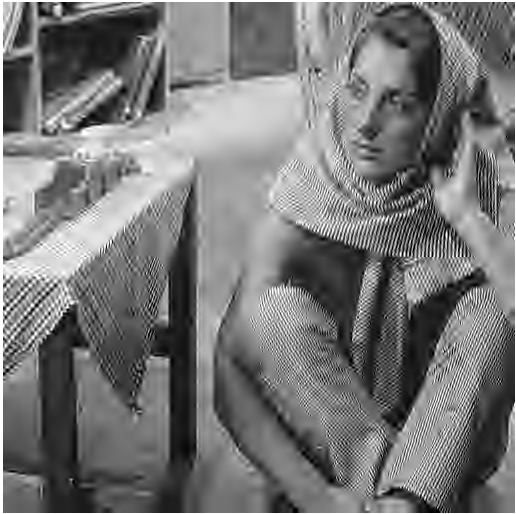}}}
	{\subfloat[2-packet reconstruction  (26.28 dB)]{\label{fig:barbara_jpeg2000_overall9packets_2packet_approx__26_28dB__0_153bpp}\includegraphics[width=0.31\textwidth]{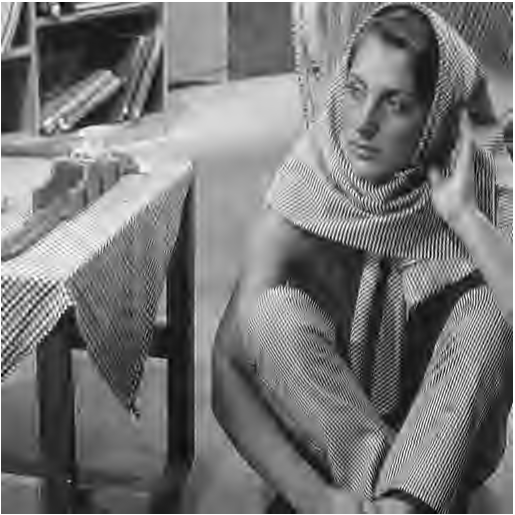}}}
	{\subfloat[3-packet reconstruction (27.47)]{\label{fig:barbara_jpeg2000_overall9packets_3packet_approx__27_47dB__0_153bpp}\includegraphics[width=0.31\textwidth]{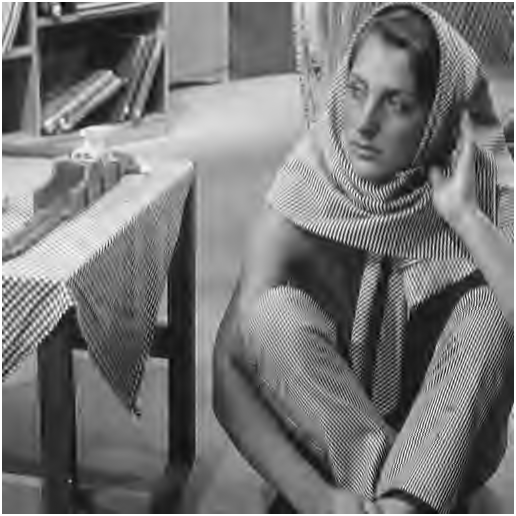}}}
	\\
	{\subfloat[4-packet reconstruct (29.10 dB)]{\label{fig:barbara_jpeg2000_overall9packets_4packet_approx__29_10dB__0_153bpp}\includegraphics[width=0.31\textwidth]{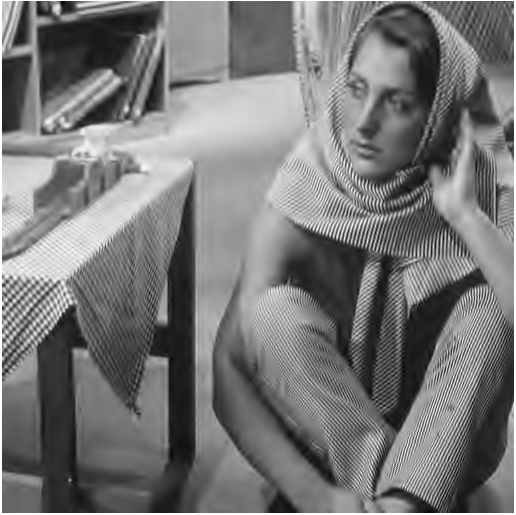}}}
	{\subfloat[5-packet reconstruction  (30.11 dB)]{\label{fig:barbara_jpeg2000_overall9packets_5packet_approx__30_11dB__0_153bpp}\includegraphics[width=0.31\textwidth]{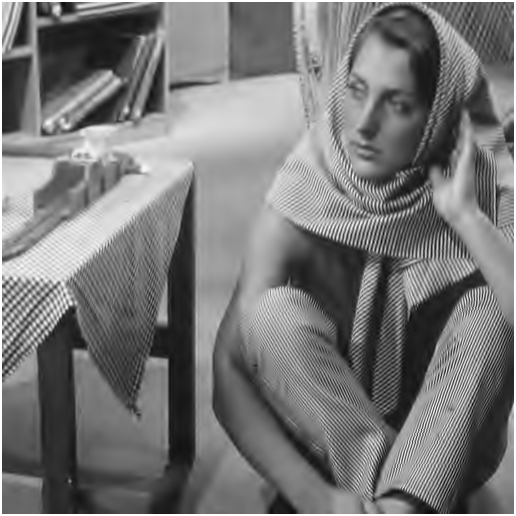}}}
	{\subfloat[6-packet reconstruction (30.50)]{\label{fig:barbara_jpeg2000_overall9packets_6packet_approx__30_50dB__0_153bpp}\includegraphics[width=0.31\textwidth]{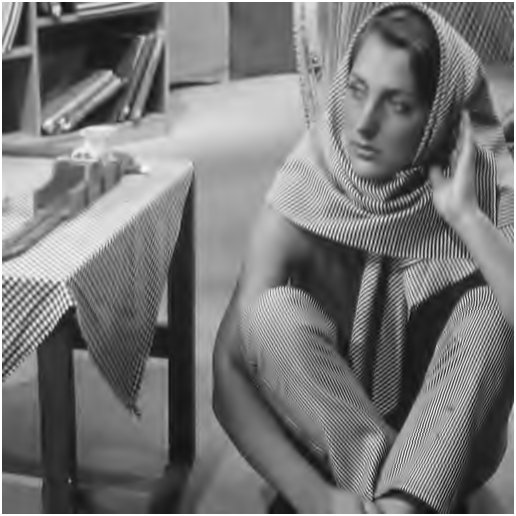}}}
	\\	
	{\subfloat[7-packet reconstruct (31.33 dB)]{\label{fig:barbara_jpeg2000_overall9packets_7packet_approx__31_33dB__0_153bpp}\includegraphics[width=0.31\textwidth]{figures/barbara_jpeg2000_overall9packets_4packet_approx__29_10dB__0_153bpp.png}}}
	{\subfloat[8-packet reconstruction  (31.89 dB)]{\label{fig:barbara_jpeg2000_overall9packets_8packet_approx__31_89dB__0_153bpp}\includegraphics[width=0.31\textwidth]{figures/barbara_jpeg2000_overall9packets_5packet_approx__30_11dB__0_153bpp.png}}}
	{\subfloat[9-packet reconstruction (32.19)]{\label{fig:barbara_jpeg2000_overall9packets_9packet_approx__32_19dB__0_153bpp}\includegraphics[width=0.31\textwidth]{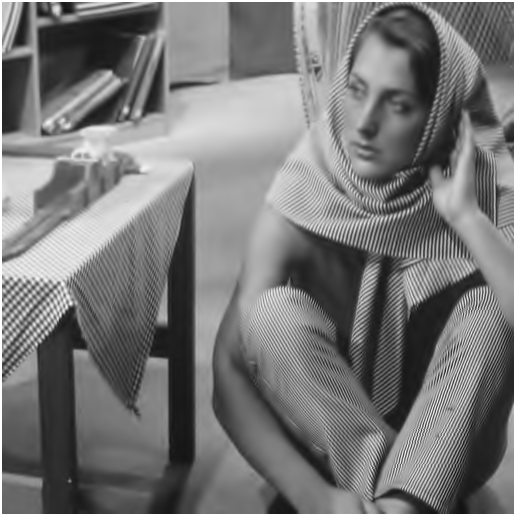}}}
	\\	
	\caption{Examples for $ m $-packet reconstructions of the 'Barbara' image using multiple packets from the set of 9 holographic representations.
		Demonstration of $ m $-packet reconstructions obtained from a set of 9 holographic packets optimized by the proposed framework for a 9-packet reconstruction.  The utilized compression is JPEG2000 at a compression ratio of 1:50.} 
	\label{Fig:Experiments - barbara - optimized for 9 holographic representations}
\end{figure*}

In addition, we also examine the case where the complete set of representations includes 9 packets. In this case, the shifts are based on offsets of the upper-left coordinate of the image by $ (3\Delta_x,3\Delta_y)$ pixels for all $\Delta_x,\Delta_y \in \left\lbrace 0,1,2 \right\rbrace$. The formulas for setting the parameters are as in the first setting described above, except for the $ \lambda $ parameter in the 2-packet optimization mode that is now set to $ K^2 $. 
The comparison presented in Fig. \ref{Fig:Experiments - 9 holographic representations -  PSNR-number of packets curves} and Table \ref{table:Experiments - blur - Average PSNR and Bit-Rate comparison - 9 Packets - Compression Ratio 50} demonstrates the improvements in PSNR achievable using the proposed optimization framework. In Fig. \ref{Fig:Experiments - barbara - optimized for 9 holographic representations} we visually demonstrate the progressive refinement when increasing the number of packets utilized.

\begin{table*} []
	\footnotesize
	\caption{Evaluation of Quality and Diversity in the Reconstructions From a Set of 4 Packets  (The Mean and Standard Deviation Values Refer to PSNR values in dB units): The Results are Based on JPEG2000 Compression at 1:50 Compression Ratio}
	\renewcommand{\arraystretch}{0.9}
	\label{table:Experiments - blur - Average PSNR and Bit-Rate comparison}
	\centering
	\begin{tabular}{|c|c||c|c||c|c||c|c||c|c|}
		\hline
		\multirow{2}{*}{\bfseries \shortstack{Image}}  & \multirow{2}{*}{\bfseries \shortstack{Method}}  & \multicolumn{2}{|c|}{\bfseries \shortstack{\\1 Packet}} & \multicolumn{2}{|c|}{\bfseries \shortstack{\\2 Packets}} & \multicolumn{2}{|c|}{\bfseries \shortstack{\\3 Packets}} & \multicolumn{2}{|c|}{\bfseries \shortstack{\\4 Packets}}\\
		\cline{3-10}
		& & \shortstack{\\$\meanDistortion{1}$} & \shortstack{\\$\stdDistortion{1}$} & \shortstack{\\$\meanDistortion{2}$} & \shortstack{\\$\stdDistortion{2}$} &  \shortstack{\\$\meanDistortion{3}$} & \shortstack{\\$\stdDistortion{3}$} &  \shortstack{\\$\meanDistortion{4}$} & \shortstack{\\$\stdDistortion{4}$} \\
		\hline
		\hline				                                       
		Cameraman & Exact Duplication  & \textbf{25.66} & 0 & 25.66 & 0 &  25.66 & 0 & 25.66 & 0 \\
		\cline{2-10}	
		& Baseline (Unoptimized)  & {25.55} & 0.10 & 26.41 & {0.16} & 26.74  & 0.06 & 26.92 & 0 \\
		\cline{2-10}	
		& Optimized for 2-Packet Reconstruction  & 25.07 & 0.06 & \textbf{27.04} & 0.20 & 27.95  & {0.03} & 28.50 & 0 \\
		\cline{2-10}	
		& Optimized for 4-Packet Reconstruction  & 23.01 & {0.03} & 26.24 & 0.20 & \textbf{28.23} & 0.06 & \textbf{29.73} & 0 \\
		\hline
		\hline
		House & Exact Duplication  & 31.14 & 0 & 31.14 & 0 &  31.14 & 0 & 31.14 & 0 \\
		\cline{2-10}	
		& Baseline (Unoptimized)  & \textbf{31.23} & 0.06 & 32.24 & {0.20} & 32.63  & 0.03 &  32.84 & 0 \\
		\cline{2-10}	
		& Optimized for 2-Packet Reconstruction  & 30.42 & 0.12 & \textbf{32.64} & 0.33 & 33.72  & 0.07 & 34.39 & 0 \\
		\cline{2-10}	
		& Optimized for 4-Packet Reconstruction  & 28.19 & {0.04} & 31.64 & 0.46 & \textbf{33.85} & {0.02} & \textbf{35.62} & 0 \\
		\hline
		\hline
		Lena & Exact Duplication  & 31.81 & 0 & 31.81 & 0 & 31.81  & 0 & 31.81 & 0 \\
		\cline{2-10}	
		& Baseline (Unoptimized)  & \textbf{31.86} & {0.03} & 32.86 & {0.20} & 33.24  & 0.03 & 33.45  & 0 \\
		\cline{2-10}	
		& Optimized for 2-Packet Reconstruction  & 31.25 & {0.03} & \textbf{33.35} & 0.30 & 34.35  & {0.02} & 34.95 & 0 \\
		\cline{2-10}	
		& Optimized for 4-Packet Reconstruction  & 28.75 & 0.13 & 32.22 & 0.40 & \textbf{34.45} & 0.04 & \textbf{36.25} & 0 \\		
		\hline		
		\hline
		Barbara & Exact Duplication  & \textbf{26.12} & 0 & 26.12 & 0 & 26.12  & 0 & 26.12 & 0 \\
		\cline{2-10}	
		& Baseline (Unoptimized)  & \textbf{26.12} & 0.04 & 27.29 & {0.13} & 27.76  & {0.01} & 28.02 & 0 \\
		\cline{2-10}	
		& Optimized for 2-Packet Reconstruction  & 25.31 & {0.01} & \textbf{27.70} & 0.19 &  28.91 & 0.05 &  29.67 & 0 \\
		\cline{2-10}	
		& Optimized for 4-Packet Reconstruction  & 22.51 & 0.10 & 26.30 & 0.45 & \textbf{28.96} & 0.07 & \textbf{31.39} & 0 \\
		\hline
	\end{tabular}
\end{table*}

\begin{table*} []
	\footnotesize
	\caption{Evaluation of Quality and Diversity in the Reconstructions From a Set of 4 Packets (The Mean and Standard Deviation Values Refer to PSNR values in dB units): The Results are Based on JPEG2000 Compression at 1:25 Compression Ratio}
	\renewcommand{\arraystretch}{0.9}
	\label{table:Experiments - blur - Average PSNR and Bit-Rate comparison - Compression Ratio 25}
	\centering
	\begin{tabular}{|c|c||c|c||c|c||c|c||c|c|}
		\hline
		\multirow{2}{*}{\bfseries \shortstack{Image}}  & \multirow{2}{*}{\bfseries \shortstack{Method}}  & \multicolumn{2}{|c|}{\bfseries \shortstack{\\1 Packet}} & \multicolumn{2}{|c|}{\bfseries \shortstack{\\2 Packets}} & \multicolumn{2}{|c|}{\bfseries \shortstack{\\3 Packets}} & \multicolumn{2}{|c|}{\bfseries \shortstack{\\4 Packets}}\\
		\cline{3-10}
		& & \shortstack{\\$\meanDistortion{1}$} & \shortstack{\\$\stdDistortion{1}$} & \shortstack{\\$\meanDistortion{2}$} & \shortstack{\\$\stdDistortion{2}$} &  \shortstack{\\$\meanDistortion{3}$} & \shortstack{\\$\stdDistortion{3}$} &  \shortstack{\\$\meanDistortion{4}$} & \shortstack{\\$\stdDistortion{4}$} \\
		\hline
		\hline	
				Cameraman & Exact Duplication  & \textbf{28.86} & 0 & 28.86 & 0 &  28.86 & 0 & 28.86 & 0 \\
				\cline{2-10}	
				& Baseline (Unoptimized)  & {28.84} & 0.06 & 30.01 & {0.14} & 30.47  & 0.04 & 30.73 & 0 \\
				\cline{2-10}			
				& Optimized for 2-Packet Reconstruction  & 28.34 & 0.04 & \textbf{30.35} & 0.22 & 31.29  & {0.03} & 31.85 & 0 \\
				\cline{2-10}			
				& Optimized for 4-Packet Reconstruction  & 26.03 & {0.08} & 29.54 & 0.24 & \textbf{31.85} & 0.02 & \textbf{33.74} & 0 \\
				\hline
				\hline		
				House & Exact Duplication  & 34.58 & 0 & 34.58 & 0 & 34.58  & 0 & 34.58 & 0 \\
				\cline{2-10}	
				& Baseline (Unoptimized)  & \textbf{34.59} & 0.06 & 35.54 & {0.14} & 35.91  & 0.01 &  36.10 & 0 \\
				\cline{2-10}	
				& Optimized for 2-Packet Reconstruction  & 34.20 & 0.03 & \textbf{35.85} & 0.19 & 36.58  & 0.01 & 36.99 & 0 \\	
				\cline{2-10}	
				& Optimized for 4-Packet Reconstruction  & 31.41 & {0.14} & 34.79 & 0.30 & \textbf{36.93} & {0.02} & \textbf{38.62} & 0 \\
				\hline
				\hline		
				Lena & Exact Duplication  & \textbf{35.00} & 0 & 35.00 & 0 & 35.00  & 0 & 35.00 & 0 \\
				\cline{2-10}	
				& Baseline (Unoptimized)  & \textbf{35.00} & {0.02} & 36.06 & {0.14} & 36.48  & 0.01 & 36.71  & 0 \\
				\cline{2-10}		
				& Optimized for 2-Packet Reconstruction  & 34.57 & {0.02} & \textbf{36.38} & 0.23 & 37.18  & {0.01} & 37.66 & 0 \\		
				\cline{2-10}	
				& Optimized for 4-Packet Reconstruction  & 32.45 & 0.05 & 35.60 & 0.36 & \textbf{37.49} & 0.04 & \textbf{38.87} & 0 \\		
				\hline
				\hline			                                       
		Barbara & Exact Duplication  & \textbf{29.34} & 0 & 29.34 & 0 & 29.34  & 0 & 29.34 & 0 \\
		\cline{2-10}	
		& Baseline (Unoptimized)  & {29.33} & 0.02 & 30.88 & {0.22} & 31.54  & {0.01} & 31.92 & 0 \\
		\cline{2-10}			
		& Optimized for 2-Packet Reconstruction  & 28.86 & {0.02} & \textbf{31.27} & 0.33 &  32.48 & 0.02 &  33.26 & 0 \\		
		\cline{2-10}	
		& Optimized for 4-Packet Reconstruction  & 26.58 & 0.09 & 30.33 & 0.50 & \textbf{32.91} & 0.03 & \textbf{35.22} & 0 \\
		\hline
	\end{tabular}
\end{table*}

\begin{table*} []
	\footnotesize
	\caption{Evaluation of Quality and Diversity in the Reconstructions From a Set of 9 Packets (The Mean and Standard Deviation Values Refer to PSNR values in dB units): The Results are Based on JPEG2000 Compression at 1:50 Compression Ratio.~~~This table presents the mean and standard deviation for reconstructions using 1,2,3 and 9 packets. The corresponding properties for reconstructions based on 4,5,7, and 8 packets can be coarsely examined using the curves in Fig. \ref{Fig:Experiments - 9 holographic representations -  PSNR-number of packets curves}.}
	\renewcommand{\arraystretch}{0.9}
	\label{table:Experiments - blur - Average PSNR and Bit-Rate comparison - 9 Packets - Compression Ratio 50}
	\centering
	\begin{tabular}{|c|c||c|c||c|c||c|c||c|c|}
		\hline
		\multirow{2}{*}{\bfseries \shortstack{Image}}  & \multirow{2}{*}{\bfseries \shortstack{Method}}  & \multicolumn{2}{|c|}{\bfseries \shortstack{\\1 Packet}} & \multicolumn{2}{|c|}{\bfseries \shortstack{\\2 Packets}} & \multicolumn{2}{|c|}{\bfseries \shortstack{\\3 Packets}} & \multicolumn{2}{|c|}{\bfseries \shortstack{\\9 Packets}}\\
		\cline{3-10}
		& & \shortstack{\\$\meanDistortion{1}$} & \shortstack{\\$\stdDistortion{1}$} & \shortstack{\\$\meanDistortion{2}$} & \shortstack{\\$\stdDistortion{2}$} &  \shortstack{\\$\meanDistortion{3}$} & \shortstack{\\$\stdDistortion{3}$} &  \shortstack{\\$\meanDistortion{9}$} & \shortstack{\\$\stdDistortion{9}$} \\
		\hline
		\hline				                                       
		Cameraman & Exact Duplication  & \textbf{25.66} & 0 & 25.66 & 0 &  25.66 & 0 & 25.66 & 0 \\
		\cline{2-10}
		& Baseline (Unoptimized)  & {25.57} & 0.10 & 26.43 & {0.21} & 26.74  & {0.17} & 27.19 & 0 \\
		\cline{2-10}			
		& Optimized for 2-Packet Reconstruction  & 25.40 & 0.10 & \textbf{26.73} & 0.24 & 27.26  & 0.18 & 28.09 & 0 \\
		\cline{2-10}			
		& Optimized for 9-Packet Reconstruction  & 23.55 & {0.10} & 26.15 & 0.25 & \textbf{27.53} & 0.21 & \textbf{30.51} & 0 \\
		\hline
		\hline		
		House & Exact Duplication  & 31.14 & 0 & 31.14 & 0 &  31.14 & 0 & 31.14 & 0 \\
		\cline{2-10}	
		& Baseline (Unoptimized)  & \textbf{31.29} & {0.06} & 32.30 & {0.21} & 32.67  & {0.16} &  33.22 & 0 \\
		\cline{2-10}			
		& Optimized for 2-Packet Reconstruction  & 31.06 & 0.10 & \textbf{32.46} & 0.29 & 33.03  & 0.22 & 33.97 & 0 \\
		\cline{2-10}			
		& Optimized for 9-Packet Reconstruction  & 29.21 & 0.20 & 31.81 & 0.41 & \textbf{33.18} & 0.33 & \textbf{36.13} & 0 \\
		\hline
		\hline		
		Lena & Exact Duplication  & 31.81 & 0 & 31.81 & 0 & 31.81  & 0 & 31.81 & 0 \\
		\cline{2-10}
		& Baseline (Unoptimized)  & \textbf{31.88} & {0.04} & 32.89 & {0.19} & 33.25  & {0.15} & 33.82  & 0 \\
		\cline{2-10}			
		& Optimized for 2-Packet Reconstruction  & 31.66 & 0.05 & \textbf{33.03} & 0.23 & 33.64  & 0.20 & 34.59 & 0 \\
		\cline{2-10}			
		& Optimized for 9-Packet Reconstruction  & 29.71 & 0.15 & 32.36 & 0.32 & \textbf{33.82} & 0.28 & \textbf{36.90} & 0 \\		
		\hline		
		\hline		
		Barbara & Exact Duplication  & {26.12} & 0 & 26.12 & 0 & 26.12  & 0 & 26.12 & 0 \\
		\cline{2-10}
		& Baseline (Unoptimized)  & \textbf{26.17} & {0.05} & 27.24 & {0.23} & 27.67  & {0.19} & 28.34 & 0 \\
		\cline{2-10}					
		& Optimized for 2-Packet Reconstruction  & 25.78 & {0.05} & \textbf{27.42} & 0.33 &  \textbf{28.11} & 0.27 &  29.29 & 0 \\		
		\cline{2-10}			
		& Optimized for 9-Packet Reconstruction  & 23.48 & 0.20 & 26.41 & 0.48 & 28.01 & 0.44 & \textbf{32.19} & 0 \\
		\hline
	\end{tabular}
\end{table*}

\section{Conclusion}
\label{sec:Conclusion}
In this paper we proposed a new methodology for signal and image compression, intended for systems where compressed data is often trivially duplicated in exact forms. Our idea relies on the concept of holographic representations that are equally descriptive and useful for progressive refinement of the reconstructed signal. Based on the shift-sensitivity of signal compression techniques, we developed a baseline and an ADMM-based optimized framework for the construction of binary compressed representations compatible with standard compression techniques. Our experiments clearly demonstrate the effectiveness of the proposed framework, reaching remarkable improvements in the reconstruction quality over the approach of using exact duplications. Future work can extend the proposed framework for optimizing holographic compression based on projection operators other than shifts. Moreover, the guidelines established here for optimized holographic compression can be generalized further to holographic representations using various regularization types, replacing the role of the bit-cost measures in this paper.


\ifCLASSOPTIONcaptionsoff
  \newpage
\fi



\bibliographystyle{IEEEtran}
\bibliography{IEEEabrv,holographic_compression_via_shifts_conference__refs}
%

%
%

%





\end{document}